\documentclass[fleqn,twoside]{article}
\usepackage[headings]{espcrc2}
\usepackage{graphicx}
\usepackage{graphics}
\usepackage{epsfig}
\usepackage{amssymb}
\usepackage{cite}

\newcommand{\mt}{m_t}
\newcommand{\mb}{m_b}

\newcommand{\tb}{\tan\beta}
\newcommand{\msusy}{M_{\rm SUSY}}
\newcommand{\mg}{m_{\tilde{g}}}
\newcommand{\bsg}{b\to s \gamma}

\usepackage{graphicx}
\usepackage[figuresright]{rotating}

\title{Single top-quark production by direct supersymmetric\\
flavor-changing neutral-current interactions at the LHC}

\author{Jaume Guasch\address[ECM]{HEP Group, Dept. Estructura i
Constituents de la Mat\`eria,\\ Universitat de Barcelona, Av.
Diagonal 647, E-08028 Barcelona, Catalonia, Spain}, Wolfgang
Hollik\address{Max-Planck-Institut f\"ur Physik, F\"ohringer Ring
6, D-80805, Munich, Germany}, Siannah Pe\~naranda\address{CERN TH
Division, Dept. Physics, CH-1211 Geneva 23, Switzerland},
\underline{Joan Sol\`a}\addressmark[ECM]\thanks{Contribution to
the proceedings of \textit{7th International Symposium on
Radiative Corrections},  Shonan Village, Japan, October 2-7,
2005}}

\runtitle{Single top-quark production by direct supersymmetric
FCNC interactions at the LHC} \runauthor{J. Guasch, W. Hollik, S.
Pe\~naranda, J. Sol\`a}

\begin{document}
\hyphenation{e-ner-gy bran-ching ty-pi-cal do-mi-nant}

\begin{abstract}
Production of (electrically neutral) heavy-quark pairs, such as
$t{\bar c}$ and ${\bar t}c$, is extremely suppressed in the SM. In
supersymmetric (SUSY) theories, such as the MSSM, the number of
these events can be significantly enhanced thanks (mainly) to the
FCNC couplings of gluinos. We compute the efficiency of this
mechanism for FCNC production of heavy quarks at the LHC. We find
that $\sigma (pp\rightarrow t\bar{c}+\bar{t}c)$ can reach $1\,{\rm
pb}$, and therefore one can expect up to $10^{5}$ events per
$100\,{\rm fb}^{-1}$ of integrated luminosity (with no
counterpart in the SM). Their detection would be instant evidence
of new physics, and could be a strong indication of underlying
SUSY dynamics. \vspace{1pc}
\end{abstract}

\maketitle

\section{Introduction}

Flavor-changing neutral-current (FCNC) interactions of top quarks
are among the most promising processes to deal with as a probe of
new physics. This is so because that kind of processes are (in
contrast to low-energy meson FCNC physics) extremely suppressed in
the SM. For instance, while radiative $B$-meson decays have
branching ratios of order $B(b\rightarrow s\gamma)\sim 10^{-4}$,
typical FCNC top-quark decays, such as $t\rightarrow Z c$ and
$t\rightarrow g c$, have SM branching ratios of order of at most
$10^{-13}$ and $10^{-11}$ respectively\, \cite{LorenzoEilam},
which in practice are impossible to measure. And among these FCNC
processes the rarest ones in the SM are those involving the top
quark and the Higgs boson, e.g. $B(t\rightarrow H_{\rm SM}
\,c)\sim 10^{-14}$\,\cite{Mele} and the crossed one $B(H_{\rm
SM}\rightarrow t\,\bar{c})\sim 10^{-13}$-$10^{-16}$ (depending on
the Higgs mass)\cite{BGS2}. Fortunately, when one considers the
impact of new physics (e.g. Supersymmetry or generalized Higgs
sectors) the situation may change dramatically. Indeed, as first
emphasized in the detailed work of\, \cite{GS99}, the FCNC
processes involving top quarks and Higgs bosons may constitute a
prominent door to SUSY physics in high-luminosity colliders. In
that work it was found that, in contradistinction to the SM case,
the top-quark decays into MSSM Higgs bosons
$h=h^0,H^0,A^0$\,\cite{Hunter} can be the most favored FCNC top
decays of all, with branching ratios that can reach the level of
$B(b\rightarrow s\gamma)$. This is not possible (without
fine-tuning) for the FCNC top quark decays into gauge bosons in
the MSSM, which stay typically two orders of magnitude
below\,\cite{tG}. Similarly, the maximal MSSM Higgs boson FCNC
rates into top-quark final states, e.g. $H^0,A^0\rightarrow t{\bar
c}+{\bar t}c$, can be of order $10^{-4}$\,\cite{FCNCMSSM1}, which
suggests that these decays could be a source of a sizeable number
of FCNC events $t{\bar c}$ and ${\bar t}c$ in a high-luminosity
collider. Actually, a detailed calculation of the number of these
events at the LHC has recently been reported in\,\cite{BGS4} and
confirmed this expectation, to wit: a few thousand FCNC events per
$100\,{{\rm fb}}^{-1}$ of integrated luminosity are possible.
Furthermore, a number of works have stressed the importance of
this kind of FCNC processes in more general two-Higgs-doublet
models (2HDM)\,\cite{BGS2,BGS1}, including some effects that could
appear in multiple Higgs models\,\cite{CMunoz}. Finally, let us
also mention the possibility of enhanced FCNC top-quark effects in
top-color assisted technicolor models\,\cite{topcolor}. A general,
model-independent, parametrization of new FCNC effects is
presented in\,\cite{Ferreira}.

Interestingly enough, there also exists the possibility to
produce $t{\bar c}$ and ${\bar t}c$ final states without Higgs
bosons or any other intervening particle. In this work we will
show that the FCNC gluino interactions in the MSSM can actually
be the most efficient mechanism for direct FCNC production of top
quarks.
\begin{figure}[t]
\resizebox{7cm}{!}{\includegraphics{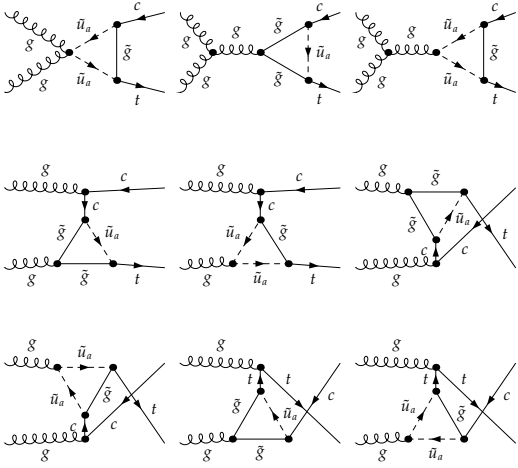}}
\caption{%
Feynman diagrams contributing to the production of $t\bar{c}$
final states at the LHC through loops of gluinos and squarks. Only
a small sample of them is shown. The FCNC interactions take place
at the vertices $\tilde{g}-\tilde{q}_i-{q}_j$ (with $i,j=c,t$),
which are proportional to $\delta_{23}$. \label{diagrams} }
\end{figure}

\section{FCNC interactions in the MSSM}

The flavor structure of the MSSM involves fermion and sfermion
mass matrices, and in general the diagonalization of the first
flavor structure does not guarantee the diagonalization of the
second. For example, the requirement of $SU(2)_L$ invariance
means that the top-left-squark mass matrix cannot be
simultaneously diagonal to the bottom-left-squark mass matrix,
and therefore these two matrices cannot be in general
simultaneously diagonal with the top-quark and bottom-quark mass
matrices. Even if we would arrange this to be so, the radiative
corrections (e.g. from the charged currents) would destroy this
arrangement. This is a sign that one cannot consistently demand
the absence of flavor-mixing interactions in the MSSM. Indeed,
even if we would ``align'' the parameters at a high energy GUT
scale, the RG running down the electroweak scale would missalign
the mass matrices\,\cite{Duncan83}. As a well-known example, let
us recall that the top-squark decay into charm quark and
neutralino ($\tilde{t}\rightarrow c\,\chi^0$) is UV-divergent in
the MSSM, unless we allow for FCNC interaction terms in the
classical Lagrangian that can absorb these
infinities\,\cite{Hikassa87}. Therefore, in general, in the MSSM
we expect terms of the form gluino--quark--squark or
neutralino--fermion--sfermion, with the quark and squark having
the same charge but belonging to different flavors. In this work
we will concentrate only on the first type of terms, which are
expected to be dominant. A detailed Lagrangian describing these
generalized SUSY--QCD interactions mediated by gluinos can be
found, e.g. in \cite{GS99}. The relevant parameters are the
flavor-mixing coefficients $\delta_{ij}$. In contrast to previous
studies\,\cite{Xinesos05}, we will allow them only in the LL part
of the $6\times 6$ sfermion mass matrices in flavor-chirality
space\,\cite{GS99}. This assumption is not only for simplicity,
but also because it is suggested by RG arguments\,\cite{Duncan83}.
Thus, if $M_{\rm LL}$ is the LL block of a sfermion mass matrix,
we define $\delta_{ij}$ ($i\neq j$) as follows: $(M_{\rm
LL})_{ij}=\delta_{ij}\,\tilde{m}_i\,\tilde{m}_j$, where
$\tilde{m}_i$ is the soft SUSY-breaking mass parameter
corresponding to the LH squark of $i$th flavor\,\cite{GS99}. We
will be mostly interested in the parameter $\delta_{23}$, because
it is the one relating the second and third generations (therefore
involving the top quark physics). $\delta_{23}$ is the less
restricted one from the phenomenological point of view. This is
because the phenomenological bounds on the various $\delta_{ij}$
(cf. \cite{Pokorsky}) are obtained from the low-energy FCNC
processes. These involve mainly the first and second generations.
Thus $\delta_{23}$ is an essentially free parameter within, say,
the natural interval $0<\delta_{23}<1$.  Actually we have two such
parameters, $\delta_{23}^{(t)}$ and $\delta_{23}^{(b)}$, for the
up-type and down-type LL squark mass matrices respectively. The
former enters the process under study whereas the latter enters
$B(\bsg)$. We will use this observable to restrict our
predictions on $t\bar{c}+\bar{t}c$ production.
\begin{figure*}[t]
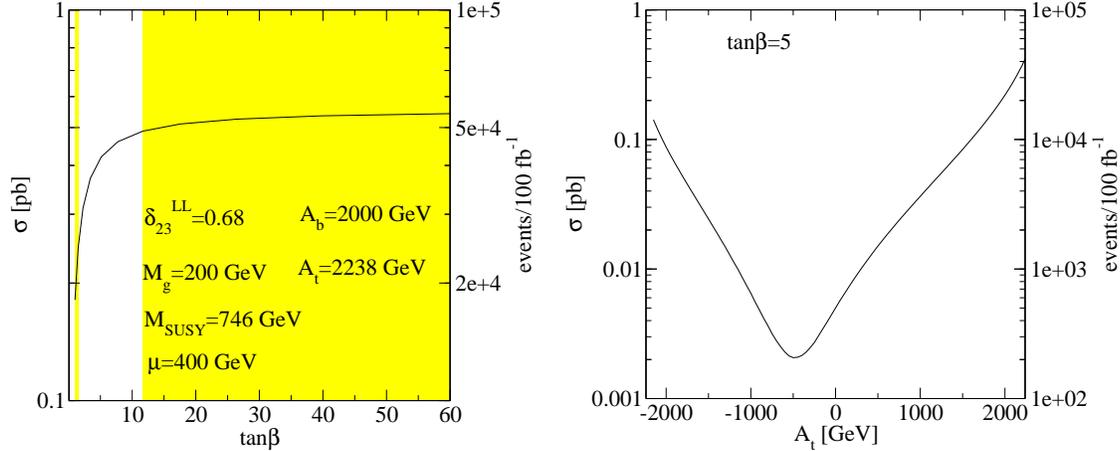

\begin{tabular}{cc}
\resizebox{!}{6cm}{\includegraphics*{max_tb.eps}}&
\resizebox{!}{6cm}{\includegraphics*{max_At.eps}}
\end{tabular}
\caption{%
$\sigma_{tc}$ (in ${\rm pb}$), Eq.(\ref{sigmatc}), and number of events
per $100\,{\rm fb}^{-1}$ of integrated luminosity at the LHC, as a
function of $\tb$ and $A_t$ for the given parameters. The shaded
region is excluded by $B_{\rm exp} (\bsg)$.\label{fig:tbAt} }
\end{figure*}

\section{Single top-quark production from glui\-no FCNC in\-te\-ractions in the MSSM}

In Fig.\,\ref{diagrams} we show some of the diagrams involved in
the direct production of the FCNC $t\bar{c}$ final states. We have
performed the calculation of the full one-loop SUSY--QCD
cross-section $\sigma_{tc}\equiv\sigma (pp\rightarrow t\bar{c})$
using standard algebraic and numerical packages for this kind of
computations\,\cite{FeynArts}. (Of course $\sigma
(pp\rightarrow t\bar{c}+\bar{t}c)=2\,\sigma_{tc}$.) The complete
formulae are very cumbersome, so to simplify the discussion it
will be sufficient to quote the general form of the cross-section:
\begin{equation}\label{sigmatc}
\sigma_{tc} \sim \left(\delta_{23}^{(t)LL}\right)^2 \, \frac{\mt^2
({A_t}-\mu/\tb)^2}{\msusy^4} \, \frac{1}{\mg^2}\,.
\end{equation}
Here $A_t$ is the trilinear top-quark coupling, $\mu$ the higgsino
mass parameter and $\msusy$ stands for the overall scale of the
squark masses -- see (\ref{eq:massmatrix}) below. The gluino mass
is denoted by $\mg$. The superscript in $\delta_{23}^{(t) {\rm LL}}$ is
to emphasize that we consider only the contributions from the LL
block of the (top-squark) mass matrix. We have performed the
computation of the above cross-section together with the branching
ratio $B(\bsg)$ in the MSSM, because only in this way can we be
sure that the region of the parameter space that we employ to
compute $\sigma_{tc}$ does respect the experimental bounds on
$B(\bsg)$. Specifically, we take $B_{\rm exp}
(\bsg)=(2.1$-$4.5)\times 10^{-4}$ at the $3\sigma$ level
\,\cite{PDB04}. Again, to ease the discussion, it suffices to
quote the MSSM formula for the branching ratio as follows\,\cite{BGS4}:
\begin{equation}\label{bsg}
B(\bsg) \sim \left(\delta_{23}^{(b){\rm LL}}\right)^2 \, \frac{\mb^2
({A_b}-\mu\tb)^2}{\msusy^4}\,.
\end{equation}
Notice that  $\delta_{23}^{(b) {\rm LL}}$ from the down-quark
mass matrix is related to the parameter $\delta_{23}^{(t) {\rm
LL}}$ in (\ref{sigmatc}) (from the up-quark mass matrix) because
the two LL blocks of these matrices are precisely related by the
CKM rotation matrix $K$ as follows: $({\cal
M}_{\tilde{u}}^2)_{LL}=K\,({\cal
M}_{\tilde{d}}^2)_{LL}\,K^{\dagger}$ \,\cite{Pokorsky}.

\section{Numerical analysis}

\begin{figure*}[t]
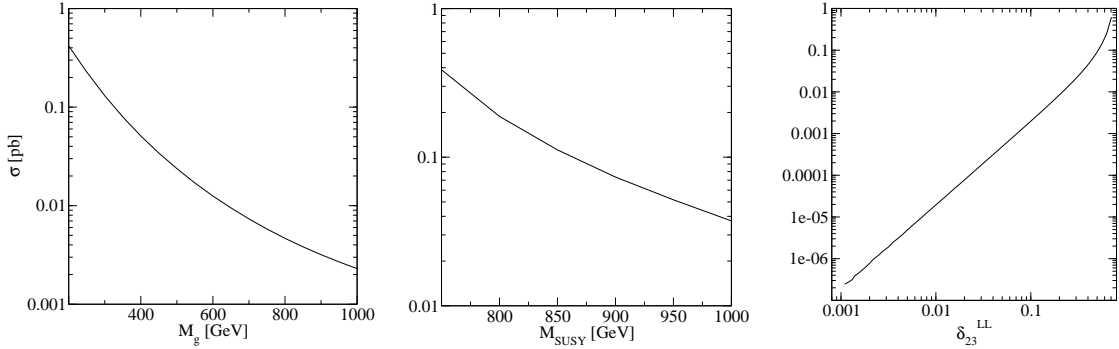

\begin{tabular}{ccc}
\resizebox{!}{4.6cm}{\includegraphics*{max_mg.eps}}&
\resizebox{!}{4.6cm}{\includegraphics*{max_msusy.eps}}&
\resizebox{!}{4.6cm}{\includegraphics*{max_d23_log.eps}}
\end{tabular}
\caption{%
As in Fig.~\ref{fig:tbAt}, but for $\sigma_{tc}$ as a function of
$\mg$, $\msusy$ and $\delta_{23}^{\rm LL}$, respectively.
\label{fig:MgMsusyd23} }
\end{figure*}

In Figs. \ref{fig:tbAt} and \ref{fig:MgMsusyd23} we present the
main results of our numerical analysis. We see that $\sigma_{tc}$
is very sensitive to $A_t$ and that it decreases with $M_{SUSY}$,
but mainly with $\mg$. As expected, it increases with
$\delta_{23}^{\rm LL}\equiv \delta_{23}^{(t) {\rm LL}}$, but it
does not reach the maximum range of this parameter. At the
maximum of $\sigma_{tc}$, it prefers $\delta_{23}^{\rm LL}=0.68$,
as we shall see below. The reason stems from the correlation of
this maximum with the $B(\bsg)$ observable. At the maximum, the
cross-section for $t\bar{c}+\bar{t}c$ production lies around
$2\sigma_{tc}\simeq 1\,{\rm pb}$, if we allow for relatively
light gluino masses $\mg=200\,$ GeV (see Fig.\,
\ref{fig:MgMsusyd23}). For higher $\mg$ the cross-section falls
down fast; at $\mg=500\,$ GeV it is already $10$ times smaller.
The total number of events per $100\,{\rm fb}^{-1}$ lies between
$10^4$-$10^5$ for this range of gluino masses. The fixed values
of the parameters in these plots lie near the values that provide
the maximum of the FCNC cross-section. The dependence on $\mu$ is
not shown, but we note that $\sigma_{tc}$ decreases by $\sim 40\%$
in the allowed range $\mu=200$-$800\,$ GeV. Values of $\mu>800\,$
GeV are forbidden by $B_{\rm exp} (\bsg)$. Large negative $\mu$
is also excluded by the experimental bound we take for the
lightest squark mass, $m_{\tilde{q}_1}\lesssim 150\,$ GeV; too
small $|\mu|\lesssim 200\,$ GeV is ruled out by the chargino mass
bound $m_{\chi^{\pm}_1}\leq 90\,$ GeV. The approximate maximum of
$\sigma_{tc}$ in parameter space has been computed using an
analytical procedure.
Let us briefly summarize the method. Define $\delta_{33}^{\rm LR}={\mt
(A_t - \mu/\tb)}/{M_{\rm SUSY}^2}$, which is involved in
(\ref{sigmatc}). Then the lines of constant $\sigma_{tc}$ are
hyperbolas in the $\delta_{33}^{\rm LR}-\delta_{23}^{\rm LL}$ plane. Next
consider the up-type squark mass matrix in the following
approximation (in particular, only the $2$nd and $3$rd squark
families are considered):
\begin{equation}
  \label{eq:massmatrix}
  {\cal M}^2_{\tilde{q}}=M_{\rm SUSY}^2\left( \begin{array}{c|ccc}
      & c_{\rm L} & t_{\rm L} & t_{\rm R} \\\hline
      c_{\rm L} & 1 & \delta_{23}^{\rm LL} & 0 \\
      t_{\rm L} & \delta_{23}^{\rm LL} & 1 & \delta_{33}^{\rm LR} \\
      t_{\rm R} & 0 &\delta_{33}^{\rm LR} &1
    \end{array}
  \right)\,\,.
\end{equation}
Upon diagonalization it is easy to see that the allowed squark
masses should lie inside the circle
$\left(\delta_{23}^{LL}\right)^2+\left(\delta_{33}^{LR}\right)^2=R^2$
whose radius is $R= 1- {m_{\tilde{q}_1}^2}/{M_{SUSY}^2}$. Notice
that $R$ increases with $M_{SUSY}$, but we impose the
(approximate) constraint $|A_t|<3\,M_{SUSY}$ to avoid
color-breaking minima. This puts a bound on $\delta_{33}^{LR}$.
Finally, looking for the point in the straight line
$\delta_{33}^{LR}=\delta_{23}^{LL}$ where the outermost hyperbola
$\sigma_{tc}={\rm const.}$ is tangent to the circle of radius $R$
in the $\delta_{33}^{\rm LR}$-$\delta_{23}^{\rm LL}$ plane, we find the
approximate maximum at 
\begin{equation}
  \delta_{23}^{LL}=\frac{\sqrt{2}}{1+\left[1+\frac29\,
  m_{\tilde{q}_1}^2/{m_t}^2\right]^{1/2}}\simeq 0.68\,.
\label{eq:maximcomb2}
\end{equation}
This is the result quoted before. The residual parameters of the
maximum easily follow. A previous analysis of this
process\,\cite{Xinesos05} did not make a systematic study of the
parameter space and did not take into account the important
restrictions imposed by $\bsg$ for $\tb>10$  (cf. Fig.\,
\ref{fig:tbAt}). That reference missed the bulk of the
contribution and tended to emphasize that the main effects stem
from the LR sector of the full mass matrix ${\cal
M}^2_{\tilde{q}}$, namely from $\delta_{ij}^{LR}$ ($i\neq j$). In
contrast, we have proved that it suffices to consider the LL
sector, which is the only one that is well motivated by
renormalization group arguments\,\cite{Duncan83}.

Finally, we note that $t\bar{c}$ final states can also be
produced at one-loop by the charged-current interactions within
the SM. We have computed this one-loop cross-section at the LHC,
with the result
$\sigma^{\rm SM}(pp\rightarrow t\bar{c}+\bar{t}c)=7.2\times
10^{-4}\,{\rm fb}\,.$
It amounts to less than one event in the entire lifetime of the
LHC. So it is pretty obvious that only the presence of new physics
could be an explanation for these events, if they are ever
detected.

\section{Discussion and conclusions}

We have computed the full one-loop SUSY--QCD cross-section for the
production of single top-quark states $t{\bar c}+{\bar t}c$ at the
LHC.  We have shown that this direct production mechanism is
substantially more efficient (typically a factor of $100$) than
the production and subsequent FCNC decay\,\cite{BGS4} of the heavy
MSSM Higgs bosons ($H^0,A^0\rightarrow t{\bar c}+{\bar t}c$). It
is important to emphasize that, if the mass generation mechanism
is associated to a fundamental Higgs sector, then the detection of
a significant number of $t{\bar c}+{\bar t}c$ states could be
interpreted as a distinctive SUSY signature. The reason for this
is that in an unconstrained 2HDM (types I and II), these FCNC
events cannot be produced at comparable rates. There is no direct
production mechanism in this case (at one-loop), and therefore a
significant $t{\bar c}+{\bar t}c$ signature could only come from
Higgs-boson decays whose efficiency, though, was shown to be
comparatively much smaller\,\cite{BGS2}, namely only a few
hundred events could be expected versus $10^5$ events that can be
achieved by direct production in the MSSM. Given the robust
signal carried by the single top-quark in the final state, these
FCNC processes could be a very helpful tool to complement the
search for SUSY physics at the LHC collider. Before closing we
point out that there are also direct SUSY--EW (electroweak) loop
diagrams (complementing the SUSY--QCD ones in
Fig.\,\ref{diagrams}), which could be important in certain
regions of the MSSM parameter space. The corresponding analysis
of these SUSY--EW effects will be presented elsewhere.

\textit{Note added.} After the present work was first submitted,
we noticed the recent reference \cite{Eilam} which addresses the
same subject.

\textbf{Acknowledgements}: JG has been supported by a {Ramon y
Cajal} contract from MEC (Spain); JG and JS in part by MEC and
FEDER under project 2004-04582-C02-01 and by DURSI Generalitat de
Catalunya under project 2005SGR00564; SP by the European Union
under contract No. MEIF-CT-2003-500030.

\newcommand{\JHEP}[3]{{JHEP} {#1} (#2)  {#3}}
\newcommand{\NPB}[3]{{ Nucl. Phys. } { B#1} (#2)  {#3}}
\newcommand{\NPPS}[3]{{\sl Nucl. Phys. Proc. Supp. } {\bf #1} (#2)  {#3}}
\newcommand{\PRD}[3]{{ Phys. Rev. } { D#1} (#2)   {#3}}
\newcommand{\PLB}[3]{{ Phys. Lett. } { #1B} (#2)  {#3}}
\newcommand{\EPJ}[3]{{ Eur. Phys. J } { C#1} (#2)  {#3}}
\newcommand{\PR}[3]{{ Phys. Rep } { #1} (#2)  {#3}}
\newcommand{\RMP}[3]{{ Rev. Mod. Phys. } { #1} (#2)  {#3}}
\newcommand{\IJMP}[3]{{ Int. J. of Mod. Phys. } { A#1} (#2)  {#3}}
\newcommand{\PRL}[3]{{ Phys. Rev. Lett. } { #1} (#2) {#3}}
\newcommand{\ZFP}[3]{{ Zeitsch. f. Physik } { C#1} (#2)  {#3}}
\newcommand{\IJMPA}[3]{{ Int. J. Mod. Phys. } { A#1} (#2) {#3}}
\newcommand{\MPLA}[3]{{ Mod. Phys. Lett. } {A#1} (19#2) {#3}}

\providecommand{\href}[2]{#2}

\end{document}